\documentclass[aps,pra,superscriptaddress,twocolumn]{revtex4}
\usepackage[T1]{fontenc}
\usepackage[utf8]{inputenc}
\usepackage{url}
\usepackage{amsmath}
\usepackage{mathtools}
\usepackage{bbold}
\usepackage{mathrsfs}
\usepackage[charter]{mathdesign}
\DeclareFontFamily{U}{dutchcal}{\skewchar\font=45 }
\DeclareFontShape{U}{dutchcal}{m}{n}{<-> s*[1.0] dutchcal-r}{}
\DeclareFontShape{U}{dutchcal}{b}{n}{<-> s*[1.0] dutchcal-b}{}
\DeclareMathAlphabet{\mathlcal}{U}{dutchcal}{m}{n}

\usepackage{graphicx}
\usepackage{epstopdf} 
\usepackage{wrapfig}
\usepackage{braket}
\usepackage{xcolor}
\usepackage[pdfstartview=FitH,colorlinks=true,citecolor=blue,linkcolor=blue,urlcolor=blue]{hyperref}
\usepackage{float}
\usepackage{mathptmx}
\usepackage{tikz}
\usepackage[vcentermath]{youngtab}
\usepackage{cancel}

\usepackage{soul}

\begin{document}

\author{Pablo Capuzzi}
\affiliation {Departamento de F\'{\i}sica, Facultad de Ciencias Exactas y Naturales, Universidad de Buenos Aires, Pabell\'on 1, Ciudad Universitaria, 1428 Buenos Aires, Argentina}
\affiliation{Instituto de F\'{\i}sica de Buenos Aires (UBA-CONICET), Pabell\'on 1, Ciudad Universitaria, 1428 Buenos Aires, Argentina}
\author{Luca Tessieri}
\affiliation{Instituto de F\'{\i}sica y Matem\'aticas, Universidad Michoacana de San Nicol\'as de Hidalgo, 58060, Morelia, Mexico}
\author{Zehra Akdeniz}
\affiliation{Faculty of Science and Letters, P\^{\i}r\^{\i} Reis University, 34940 Tuzla, Istanbul, Turkey}
\author{Anna Minguzzi} 
\affiliation{Univ. Grenoble Alpes, CNRS, LPMMC, 38000 Grenoble, France}
\author{Patrizia Vignolo}
\affiliation{Universit\'e C\^ote d’Azur, CNRS, Institut de Physique de Nice, 06200 Nice, France}
\affiliation{Institut Universitaire de France}

\title{Spin-charge separation in the quantum boomerang effect}
\begin{abstract}
  We study the localization dynamics of a SU(2) fermionic wavepacket
  launched in a (pseudo)random potential. We show that in the limit of
  strong inter-component repulsions, the total wavepacket exhibits a
  boomerang-like dynamics, returning near its initial position as
  expected for non-interacting particles, while separately each
  spin-component does not.  This spin-charge separation effect occurs
  both in the infinite repulsive limit and at finite interactions.  At
  infinite interactions, the system is integrable and thermalization
  cannot occur: the two spin-components push each other during the
  dynamics and their centers of mass stop further from each other than
  their initial position.  At finite interactions, integrability is
  broken, the two spin-components oscillate and mix, with their
  center-of-mass positions converging very slowly to the center of
  mass of the whole system.  This is a signature that the final
  localized state is a fully spin-mixed thermalized state.
 \end{abstract}

\maketitle
\section{Introduction}

In disordered quantum systems it has been shown that a wavepacket
launched with some initial velocity may return to its initial position
and stop there. This phenomenon, known as the quantum boomerang effect
(QBE) \cite{PratPRA2019,TessieriPRA2021,Sajjad2022}, occurs in the
Anderson localization (AL) limit when interactions are completely
negligible and disorder completely freezes the dynamics of the
wavepacket \cite{Anderson58,Abrahams79}.  The QBE is found not only in
real space, but also in momentum space. In this latter case, the
(pseudo)-disorder in momentum space is introduced by kicking
periodically the wavepacket. Indeed, the first experimental evidence
of the QBE was obtained in a quantum kicked-rotor experiment
\cite{Sajjad2022}.  These measurements have confirmed the theoretical
predictions \cite{PratPRA2019,TessieriPRA2021} and elucidated the
crucial role of the time-reversal symmetry in determining the presence
or absence of the QBE.  In fact, the occurrence of the QBE requires
not only the system to be in the AL regime, but also that some
symmetries of the Hamiltonian and of the initial state of the
wavepacket are fulfilled
\cite{Noronha2022,Noronha-2-2022,Janarek2022}.  When localization
takes place in momentum space, as in the case of the kicked rotor, it
is the time-reversal symmetry that is crucial for the wavepacket to
come back to its initial position. On the other hand, for systems that
localize in real space, like the Anderson model, it is the space-time
reversal symmetry that regulates the dynamics of the quantum boomerang
\cite{Noronha2022,Janarek2022}.  When this symmetry is broken, the
wavepacket --after having been launched-- stops somewhere but not
necessarily in its initial position.

Like Anderson localization, QBE is a phenomenon that is expected for
non-interacting systems, both bosonic and fermionic.  It has been
shown that weak interactions between particles partially destroy QBE:
the center of mass of the wavepacket makes a U-turn, but without
coming back to its initial position \cite{JanarekPRA2020}.  This also
happens for a one-dimensional (1D) strongly interacting Bose gas,
which can be mapped to a weakly interacting Fermi gas
\cite{Janarek2023}: interactions partially destroy interference
effects and thus the QBE. However, in the limit where the interactions
are infinitely repulsive, namely in the Tonks-Girardeau regime, the
QBE holds since the system can be mapped onto free fermions
\cite{Janarek2023}.

\begin{figure}
  \begin{center}
      \includegraphics[width=0.7\linewidth]{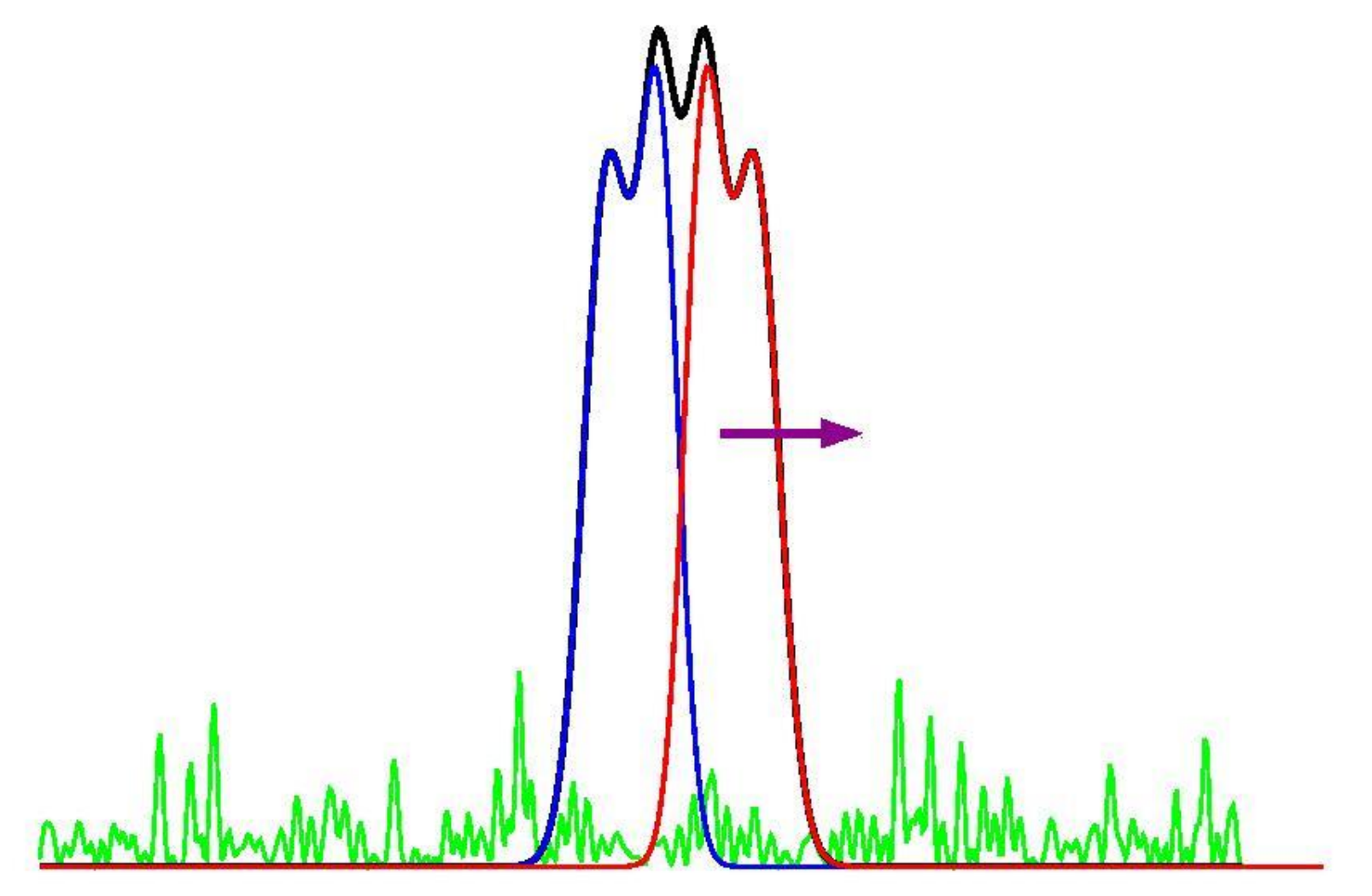}
  \end{center}
  \caption{\label{figexp}Schematic representation of the system studied in this work: a spatially spin-demixed initial state is realised with an initial non-vanishing momentum in a disorder potential.}
  \end{figure}

In this paper, we investigate the quantum boomerang dynamics of a
strongly repulsive two-component Fermi gas (see
Fig. \ref{figexp}). The underlying idea is to explore a system where
different spin configurations are available and study if the QBE, or
its failure, can bring information about the thermalization of the
system and, possibly, whether its final state is many-body localized
\cite{Basko2006,Zakrzewski2018,Kohlert2019}.

A two-component Fermi gas with strong, repulsive inter-component
contact interactions can be mapped onto an effective spin-chain
Hamiltonian where the spins exchange when particles of different
spin-components collide \cite{deuretzbacher_quantum_2014}.
  The dynamics of such a system starting from an initially
  spin-demixed configuration have been studied in the absence of
  disorder \cite{Yang2015,Yang2016,pecci2022}, highlighting the
  short-time superdiffusive dynamics of the magnetization interface
  for two spin-components happens in a similar way in spin chains
  \cite{Ljubotina2019,Ilievski2019,Denardis2019,wei2021quantum}.  It
  has also been shown that such a system relaxes towards the
  microcanonical ensemble during a time interval that increases with
  the number of particles, integrability being broken by the presence
  of an external potential at finite interactions \cite{pecci2022}.

  The presence of the disorder localizes the whole wavepacket as well
  as the two spin components. The size of the wavepacket for each
  spin-component is initially half of the total size, but very rapidly
  the two spin-components mix so that their size quickly reaches that
  of the total wavepacket.  Simultaneously, the center of mass of the
  whole system does a U-turn and stops at its initial position, as if
  the system were non-interacting.  Instead, the center-of-mass of
  each spin-component does not return to its initial position but, in
  the presence of finite interactions, converges towards the position
  of the center-of-mass of the whole system, with a damped oscillating
  dynamics much slower than that of the whole density.  This
  separation of time scales is due to the fact that the spin dynamics
  is governed by the inverse of the interaction strength, which is
  very large in our model.  We therefore observe a charge-spin
  separation in the QBE.

  Moreover, at finite interactions, where spin mixing occurs, we find
  that the final localized spin densities are those expected in the
  microcanonical ensemble. Whereas, in the limit of infinite
  interaction strength, where the center of mass of two spin-component
  stop further apart from their initial positions, the final localized
  state is very different from that expected in the microcanonical
  ensemble because integrability prevents thermalization.

  The manuscript is organized as follows.  The physical model is
  introduced in Sec. \ref{sec-model}, where we remind how the system
  can be mapped onto a spin-chain model and the dynamics can be
  exactly solved in the strong-interacting limit.  Then, the boomerang
  experiment is described in Sec. \ref{sec-exp}, where we detail a
  proposal for an experimental protocol and analyze the results,
  discussing the role of the interactions and of the symmetries.  The
  thermalization issue is discussed in Sec. \ref{sec-therm}.
  Concluding remarks are given in Sec. \ref{sec-concl}.

\section{The model}
\label{sec-model}
We consider a SU(2) fermionic mixture with
$N_\uparrow (N_\downarrow) = N/2$ fermions in the spin-up (spin-down)
component.  Each fermion is subject to an external potential $V(x)$
and interacts with fermions of the other spin component via a
repulsive contact potential of strength $g$. Thus, the many-body
Hamiltonian reads
\begin{equation}
\mathcal{H}\!=\!\mathcal{H}_0\!+\!\!\sum_{i=1}^{N_{\uparrow}}\!\sum_{j=N_\uparrow\!+1}^N\!\!\!g\delta(x_{i}-x_{j})
\end{equation}
with
\begin{equation}
\mathcal{H}_0\!=\!\sum_{i=1}^N\left[-\dfrac{\hbar^2}{2m}\dfrac{\partial^2}{\partial x^2_{i}}+V(x_{i})\right]\!\!.
 \end{equation} 
 Close to the fermionized regime, where interactions are so large that
 they play the role of a Pauli principle between fermions belonging to
 different spin components, the many-body wavefunction can be written
 as~\cite{volosniev_strongly_2014}
 \begin{equation}
    \Psi(X)=\sum_{P\in S_N}a_P\theta_P(X)\Psi_A(X)
    \label{psi}
 \end{equation}
 where the summation is performed over all $P$ permutations of $N$
 elements, $S_N$. The vector
 $X = (x_{1,\sigma_1},..., x_{N,\sigma_N})$ includes particle
 coordinates $x_i$ and spin indices $\sigma_i$, $\Psi_A(X;t)$ is the
 zero-temperature solution for spinless fermions obeying the
 non-interacting Hamiltonian $\mathcal{H}_0$, and $\theta_P(X)$ is the
 generalized Heaviside function, which is equal to 1 in the coordinate
 sector $x_{P(1),\sigma_{P(1)}}<\dots< x_{P(N),\sigma_{P(N)}}$ and
 zero otherwise.  The coefficients $a_P$ are determined by minimizing
 the energy~\cite{volosniev_strongly_2014}:
\begin{equation}
  E=E_{\infty}+\dfrac{1}{g}\left(\dfrac{\partial E}{\partial {g^{-1}}}\right)_{1/g\rightarrow 0}=E_{\infty}-\dfrac{\mathcal{C}}{g},
\end{equation}
with $\mathcal{C}=-(\partial E/\partial g^{-1})_{1/g\rightarrow 0}$
being the Tan's contact up to a dimensional constant.  This is
equivalent to solving the eigenvalue problem of the effective
Hamiltonian
\begin{equation}
\mathcal{H}_{\text{eff}}= \left.\mathcal{H}\right|_{1/g\ll1}=E_\infty \hat{1}+H_S
\end{equation}
obtained by expanding $\mathcal{H}$ on the $\{\phi_n\}$ snippet basis,
namely the basis of all particle sectors obtained by global
permutations modulo the permutations of identical fermions with the
same spin \cite{Decamp2016}.  Furthermore, it has been shown that
$H_S$ is equivalent to a spin chain Hamiltonian in position particle
space~\cite{deuretzbacher_quantum_2014}
\begin{equation}
 H_S=\sum_{j=1}^{N-1}\left( -J_j\hat{1}+J_j\hat P_{j,j+1}\right)
\label{spin_chain}
\end{equation}
where $\hat P_{j,j'}=(\vec \sigma^{(j)}\vec\sigma^{(j')}+1)/2$ is the
permutation operator and
$\vec\sigma^{(j)} =
(\sigma_{x}^{(j)},\sigma_{y}^{(j)},\sigma_{z}^{(j)})$ are the Pauli
matrices.  The hopping terms $J_i$ in Eq.~(\ref{spin_chain}) can be
written as
\begin{equation}
    J_i = \frac{N!}{g} \int_{-\infty}^{\infty} dX ~\delta(x_i - x_{i+1})  \theta_{\text{id}}(X) \Bigl\lvert \frac{\partial \Psi_A}{\partial x_i}\Big\lvert^2.
\label{eq:Ji}
\end{equation}

\subsection{The dynamics close to the fermionized regime}
In an out-of-equilibrium situation, when the free-fermion part of the
wavefunction $\Psi_A$ is time-dependent, the $J_j$ terms~(\ref{eq:Ji})
change in time.  Therefore, to obtain $\Psi(X, \bar{t}+dt)$ starting
from $\Psi(X,\bar{t})$ we proceed as follows~\cite{Barfknecht2019}.
We start by finding $J_i(\bar{t})$ to determine the spin-chain
Hamiltonian at a time $\bar{t}$.  By diagonalizing $H_S(\bar{t})$ we
obtain the eigenvectors $a_P^{(j)}(\bar{t})$ and their corresponding
eigenvalues $\mathcal{E}_j(\bar{t})$.  Expanding the coefficients
$a_{P}$ of Eq.~(\ref{psi}) in this basis gives the identity
$a_P(\bar{t})=\sum_j\alpha_j(\bar{t})a_P^{(j)}(\bar{t})$ and makes it
possible to compute the coefficients at a time $\bar{t} + dt$ as
  \begin{equation}
  a_P(\bar{t}+dt)=\sum_j\alpha_j(\bar{t})e^{-i\mathcal{E}_j(\bar{t})dt/\hbar}a_P^{(j)}(\bar{t}).
  \label{evolved_a}
  \end{equation}
  Once the evolved coefficients~(\ref{evolved_a}) are known, the
  many-body wavefunction at a time $\bar{t}+dt$ can be written as
\begin{equation}  
    \Psi(X;\bar{t}+dt)=\sum_{P\in S_N}a_P(\bar{t}+dt)\theta_P(X)\Psi_A(X;\bar{t}+dt) .
\end{equation}
For this approach to work, the time steps need to fulfil the condition
$dt\ll \hbar/|J_j(\bar{t}+dt)-J_j(\bar{t})|$ for any ${\bar{t}}$ and
any $j$.  Once the many-body wavefunction is calculated, we can
compute the spin densities at each time
\begin{equation}
    \rho_{\uparrow,\downarrow}(x,t)=\sum_{i}\delta^{\uparrow,\downarrow}_{\sigma_i}\sum_{P\in S_n}|[a_P(t)]_i|^2\rho^i(x,t)
\label{eq:rhodef}
\end{equation}
where
\begin{displaymath}
\rho^i(x,t)=\int_{x_1< \dots<x_{i-1}<x<x_{i+1}\dots<x_N} \!\!\!\!\!\!\!\!\!\!\!\!\!dX \delta(x-x_i)|\Psi_A (X,t)|^2
\end{displaymath}
is the density in the sector
$x_1<\dots<x_{i-1}<x<x_{i+1}\dots<x_N$,
while the total density is $\rho(x,t)=\sum_{i=1}^N\rho_i(x,t)=\rho_{\uparrow}(x,t)+\rho_\downarrow(x,t)$.

\subsection{The dynamics in the presence of disorder}
We now focus our attention to the case of a wavepacket of fermions
initially prepared at equilibrium in a harmonic trap of frequency
$\omega$, which is then released with an imprinted initial momentum
$\hbar k_0$ on each fermion and propagates in the pseudorandom
potential
\begin{equation}
  V_{\text{dis}}(x)=\mathcal{W}\sin(\sqrt{5}\pi(x+i_cL/2))^3/(a_{ho}/10)^3)
  \label{pseudorand}
\end{equation}
where $\mathcal{W}$ is the potential amplitude,
$a_{\text{ho}}=\sqrt{\hbar/(m\omega)}$ the typical harmonic potential
length scale, $L$ the size of the system, and $i_c$ an integer index
that counts the pseudo-disorder configurations.  It has already been
shown that the potential (\ref{pseudorand}), defined on a lattice,
induces both Anderson localisation~\cite{Griniasty1988} and the
boomerang effect~\cite{TessieriPRA2021} as a truly random potential.
We choose the potential~(\ref{pseudorand}), rather than a truly random
one, because of its relevance to cold-atom experiments, in which
pseudo-random potentials can be realized with appropriate laser
configurations~\cite{Schreiber2015}. We would like to stress, however,
that we have verified that the results described in this paper are
still valid if the pseudo-random potential~(\ref{pseudorand}) is
replaced with a random potential of adequate
strength~\cite{TessieriPRA2021}.  The potential~(\ref{pseudorand}) has
zero average $\overline{V_{\text{dis}}(x)}=0$ and is delta-correlated
$\overline{V_{\text{dis}}(x)V_{\text{dis}}(x')}=\gamma\delta(x-x')$. The
values obtained from Eq.~(\ref{pseudorand}) by keeping $x$ fixed and
varying $i_{c}$ have a uniform probability distribution function
(PDF).  Here and in the rest of this paper, we use the symbol
$\overline{(\cdots)}$ to denote the average over a sequence of
pseudo-disorder configurations (we consider different $i_c$ values in
Eq. (\ref{pseudorand})).

The disorder strength $\gamma$ determines the mean-free path $\ell$
and the mean-free time $\tau$ for a non-interacting system.  Indeed,
we remind that for a wavepacket with a momentum $\hbar k_0$ in the
Born approximation, one has~\cite{PratPRA2019}
\begin{equation}
\ell=\dfrac{\hbar^4k_0^2}{2m^2\gamma},\,\,\,\,\,\,{\rm and}\,\,\,\,\,\,
\tau=\dfrac{\hbar^3k_0}{2m\gamma}.
\end{equation}

Here and in the following, we have fixed
$\gamma=0.86\cdot 10^3\hbar^2\omega^2a_{ho}$, and $k_0=50/a_{ho}$,
that imply $\ell=1.45 a_{ho}$ and $\tau=0.029\omega^{-1}$.  We remark
that, in order to use such expressions for the case of $N$
non-interacting fermions, $(\hbar k_0)^2/(2m)$ has to be much larger
than the energy of the highest occupied orbital~\cite{Janarek2023},
namely the Fermi energy of the system.

\begin{figure}
\includegraphics[width=\columnwidth,clip=true]{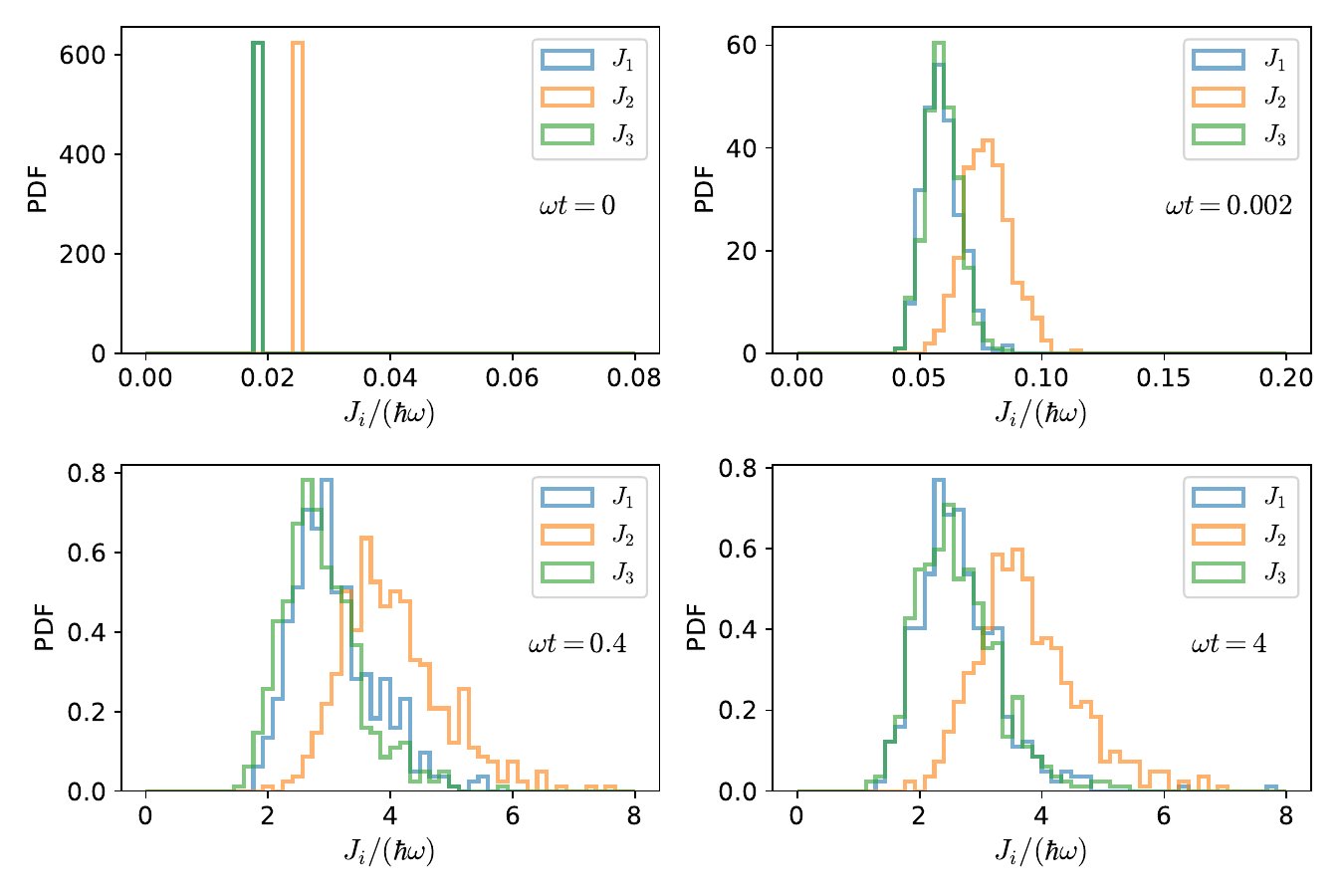}
\caption{\label{fig:Jidis}Evolution of PDF of the $J_i$ terms for the
  case of an initial deterministic wavepacket of 2+2 fermions,
  initially prepared in a harmonic trap of frequency $\omega$,
  launched with an initial momentum $\hbar k_0$ in the presence of a
  pseudorandom potential~(\ref{pseudorand}). The different panels
  correspond to times: 0, 0.002, 0.4, 4 in units of 1/$\omega$.}
\end{figure}

\begin{figure}
  \includegraphics[width=0.8\columnwidth,clip=true]{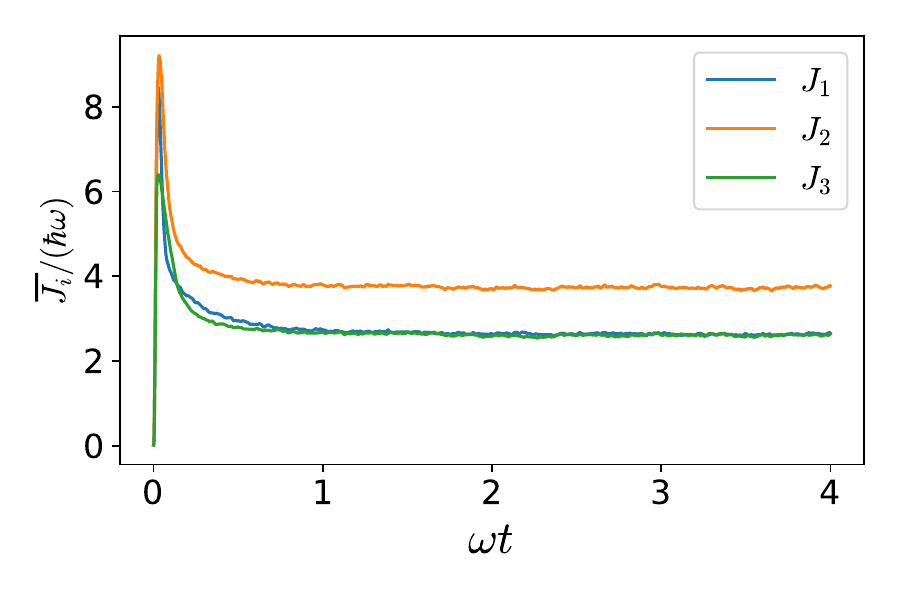}
  \includegraphics[width=0.8\columnwidth,clip=true]{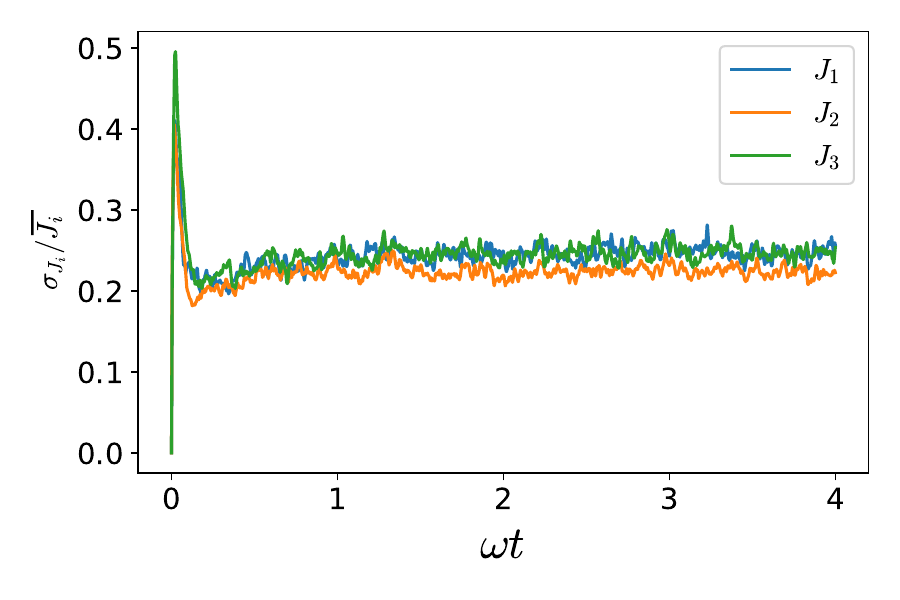}
  \caption{\label{fig:Jav}Average hopping terms $\overline{J_i}$, for
    the case $g=100\hbar\omega a_{\rm ho}$, (top panel) and relative
    standard deviation~(\ref{rescaled_stddev}) of the hopping terms
    $J_i$ (bottom panel, as functions of time, for the same
    experimental protocol as for Fig. \ref{fig:Jidis} and described in
    Sec. \ref{sec-exp}.}
\end{figure}

The presence of disorder significantly influences the time evolution
of the hopping terms $J_i$.  In an experiment in which the wavepacket
is first prepared separately in a deterministic initial condition and
then left to evolve in a disordered potential, the $J_i$'s at time
$t = 0$ are determined by the initial state and their PDFs are Dirac
deltas (as shown in the first panel of Fig.~\ref{fig:Jidis}).  Because
the fermions move in a (pseudo)-random potential, the hopping terms
become stochastic variables, with disorder entering via the time
evolution of $\Psi_A$.  As a consequence, at each time $t$, the
statistical properties of the $J_{i}$ terms must be described with a
PDF $P(J_{i})$.  The time evolution of the corresponding marginal PDFs
is depicted in Fig.~\ref{fig:Jidis}, which shows that initially the
PDFs broaden and drift towards higher values of $J_i$, eventually
reaching a stable asymptotic forms at longer times.  These qualitative
features are evinced by the time evolution of the average values of
the $J_{i}$'s and of their rescaled standard deviations
\begin{equation}
    \frac{\sigma_{J_i}}{\bar{J_{i}}} = \frac{\sqrt{ \bar{J_{i}^{2}} -
    \bar{J_{i}}^{2}}}{\bar{J_{i}}} .
    \label{rescaled_stddev}
\end{equation}
as shown in Fig.~\ref{fig:Jav}. 

We can estimate the dependence of the mean-free path $\ell_j$ and of
the mean-free time $\tau_j$ on $\bar{J_{i}}$ and $\sigma_{J_i}$ for
the spin dynamics by using the expression for the localization length
for a lattice system with random off-diagonal disorder, derived in the
Born approximation \cite{TessieriPRA2021}.  This would give
\begin{equation}
    \ell_j\propto\dfrac{\bar{J_{i}}^2}{\sigma_{J_i}^2},\quad
\text{and}
\quad
    \tau_j\propto\dfrac{\bar{J_{i}}}{\sigma_{J_i}^2}.
    \label{tauj}
\end{equation}
Equations (\ref{tauj}) predict that $\ell_j$ does not depend on the
strength of the interactions $g$, whereas $\tau_j$ increases with it
($\tau_j\propto g$), which is in agreement with the results presented
in the next section.

Finally, let's discuss the stochastic properties of the $J_i(t)$'s.
They are ultimately shaped by the random potential
$V_{\text{dis}}(x)$, but, unfortunately, the determinantal form of
$\Psi_A$ makes highly non-trivial to establish a link between the
randomness of the potential $V_{\text{dis}}(x)$ and the stochastic
features of the $J_{i}$ terms.  As shown by Eq.~(\ref{eq:Ji}), the
magnitude of the $J_i$'s is proportional to the sharpness of the
cusps, namely, the slope of the wavefunction $\Psi_{A}(X)$ when two
particles approach each other.  Therefore, we expect that their
average values should strongly depend on the specific features of the
experimental protocol.

In the present case, we consider a wavepacket launched with an initial
momentum in a disordered landscape. Our numerical analysis has shown
that the marginal PDFs of the $J_{i}$ terms tend to asymptotic forms
characterized by a skewed shape and fat tails.

\section{The boomerang dynamics}
\label{sec-exp}
We propose in this section an experimental protocol to observe QBE.
We consider an initially spatial phase-separated SU(2) fermionic
mixture with $N_{\uparrow}=N_{\downarrow}$ where the up spins are on
the left (L), and down spins on the right (R), both trapped in a
harmonic potential $V_{\text{ho}}=m\omega^2x_i^2/2$.  At $t=0$ we
release the fermions in a disorder potential $V_{\text{dis}}(x_i)$ by
kicking them towards the right with an initial momentum $\hbar k_0$
and switching off the harmonic potential.

We study the dynamics of the center-of-mass $\bar x$ and the width $w$ of the total disordered-averaged density profile
\begin{equation}
\bar x=\int \bar\rho(x) x dx/N,
\end{equation}
\begin{equation}
w=\left(\int \bar\rho(x) (x-\bar x)^2 dx/N\right)^{1/2},
\end{equation}
and we do the same for each spin component
\begin{equation} 
\bar{x}_{\uparrow,\downarrow}=\int\bar\rho_{\uparrow,\downarrow}(x) x dx/N_{\uparrow,\downarrow},
\end{equation}
\begin{equation}
w_{\uparrow,\downarrow}=\left(\int \bar\rho_{\uparrow,\downarrow}(x) (x-\bar x_{\uparrow,\downarrow})^2 dx/N_{\uparrow,\downarrow}\right)^{1/2}.
\end{equation}

Remark that, to simulate such an experimental protocol in the
procedure detailed in Sec. \ref{sec-model}, the time discretization
needs to verify $dt <\tau$.
\subsection{Spin-charge separation in the boomerang dynamics}
\begin{figure}
 \begin{center}
 \includegraphics[width=\linewidth]{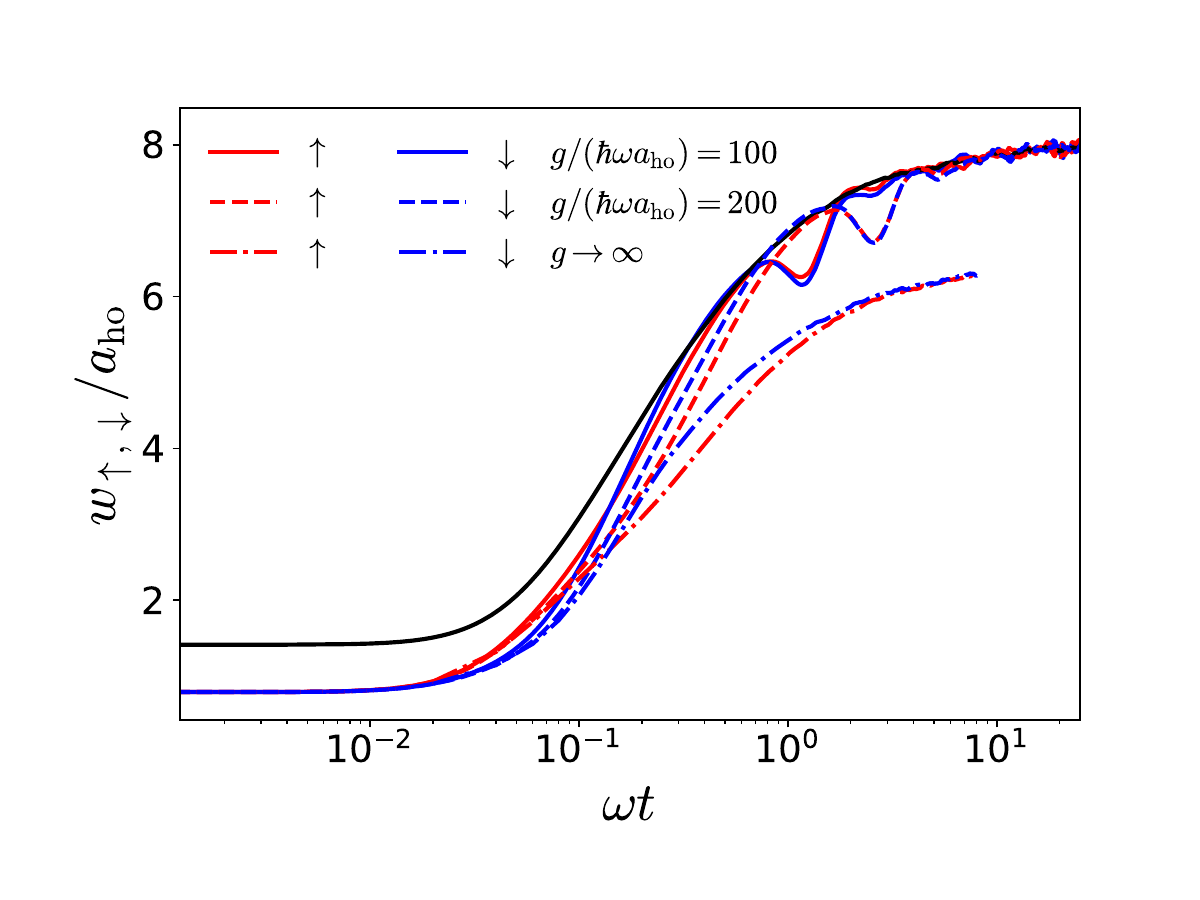}\\
 \includegraphics[width=\linewidth]{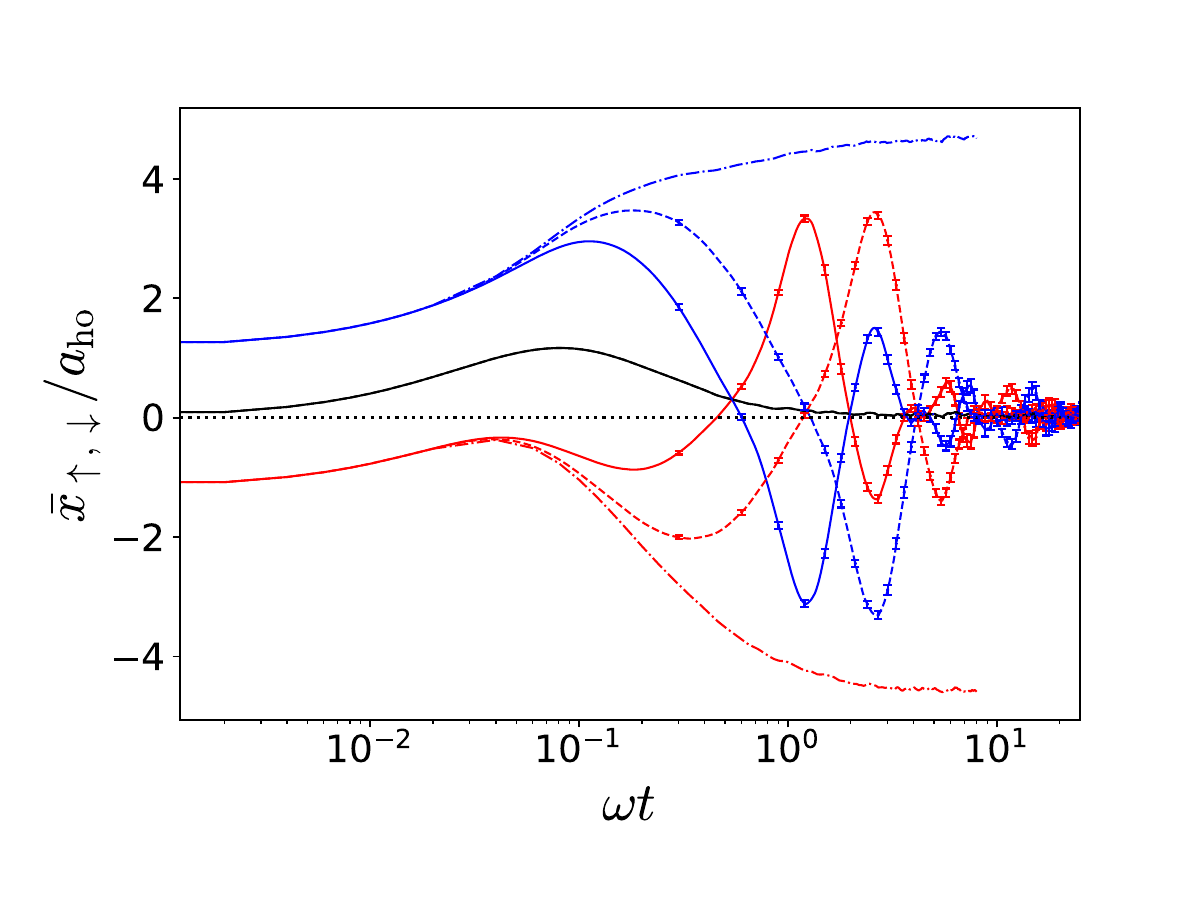}
 \end{center}
 \caption{\label{fig:boom} Disorder-averaged widths $w$ (top panel) and
   center-of-mass positions (bottom panel) $\bar x$ of the whole
   system (black curve) and of each spin component (red and blue
   lines) as functions of time (in logscale), for different values of
   the interaction strengths: $g/(\hbar\omega a_{ho})=100$ (continuous
   lines), $g/(\hbar\omega a_{ho})=200$ (dashed lines) and
   $g\rightarrow\infty$ (dot-dashed lines).  Here we used the
   following parameters: $k_0a_{ho}=50$,
   $\gamma=0.86\cdot 10^3 \hbar^2\omega^2 a_{ho}$, $N=4$ and we
   averaged over 512 configurations.}
\end{figure}

We observe (see Fig. \ref{fig:boom}) that, in the fermionized regime
($g\rightarrow\infty)$, the two components never mix, and they
localize independently, repelling each other. Indeed, in this case the
hopping terms $J_i$ vanish and
\begin{equation}
    \rho^{\infty}_\uparrow(x,t)=\rho^L(x,t)=\sum_{i=1}^{N_\uparrow} \rho^i(x,t)
\end{equation}
and
\begin{equation}
    \rho^{\infty}_\downarrow(x,t)=\rho^R(x,t)=\sum_{i=N_\uparrow+1}^N \rho^i(x,t).
\end{equation}
Moreover, the width of each component is half of the width of the
total density, and we observe that the final position of the
center-of-mass of each spin-component is
$\bar x_{\uparrow,\downarrow}^{\infty}(t\rightarrow\infty)=\pm w/2$.

The dynamical behavior is considerably different when interactions are
large but finite: in this case the two spin components undergo mixing
during the dynamics, as happens in the absence of the disorder
\cite{pecci2022} (see Fig. \ref{fig:boom}).  Each spin component is
localized by the disordered hoppings, reaching a final width that is
the same of the two components (top panel in Fig. \ref{fig:boom}).
Initially, each component moves away over a distance $\ell_{max}$, and
then it comes back towards the initial position of the center-of-mass
of the whole system, performing damped oscillations, around this
position. These oscillations are governed by the frequency spectrum of
the spin-chain Hamiltonian (\ref{spin_chain}), which yields the lowest
non-zero frequency
$\omega^\star= \left(J_1+J_2-\sqrt{J_1^2+J_2^2}\right)/\hbar$, where
we have used that at long times $J_1=J_3$. Remark that the turning
point $\ell_{max}$ is the same for the cases
$g/(\hbar\omega a _{\rm ho}))=100$ and 200, and that the time
evolution scales with $g$, namely
$\bar x_{\uparrow,\downarrow}(2t)_{2g}=\bar
x_{\uparrow,\downarrow}(t)_{g}$, as predicted by Eqs. (\ref{tauj}).
The localization dynamics occurs on a much longer timescale with
respect to that of the total density.  In particular, we observe that
at long times one has
\begin{equation}
    \bar{x}_{\uparrow,\downarrow}=\dfrac{1}{2}(\bar x_\uparrow^\infty+\bar x_\downarrow^\infty)
    \pm\dfrac{1}{2}(\bar x_\uparrow^\infty-\bar x_\downarrow^\infty)\mathcal{F}(t),
\end{equation}
with $\mathcal{F}(t)$ being a function that depends on the average of
spin weights only,
$|a_P(t)_i|^2\delta_{\sigma_i}^{\uparrow,\downarrow}$. Indeed, the
time scales of the density dynamics and of the spin dynamics being
very different (the first being determined by the disorder strength
$\gamma$ and the second by the hoppings $J_i$ that are proportional to
$\rho^3/g$ \cite{Matveev2004,deuretzbacher_momentum_2016}), the
average over configurations splits into two independent parts as if
the two dynamical processes were uncorrelated (see Appendix
\ref{app1}).

The center-of-mass of the total density exhibits the boomerang
dynamics: the whole wavepacket moves away over a distance $\sim\ell$
and then comes back to its initial position, while at the same time
the wavepacket localizes.  However, each spin-component individually
does not come back to its initial position, but its center-of-mass
position at long times coincides with the center-of-mass of the whole
system (bottom panel in Fig. \ref{fig:boom}).

Remark that the behaviour of the total density does not depend on the
value of the interaction strength in the regime we are analysing
($1/g\ll1$), so that the black curves in Fig. \ref{fig:boom} concern the
three cases analyzed ($g/(\hbar\omega a_{ho})=100, 200, \infty$).
We thus observe a spin-charge separation with respect to the boomerang
dynamics, both for finite and infinite values of the interaction
strength. This result is in agreement with the prediction of
spin-charge separation in the localization dynamics for a disordered
chain of spin-1/2 fermions \cite{Zakrzewski2018}.

\subsection{Role of the interactions and symmetries}

A straightforward interpretation of our results is the following one.
In the strongly interacting mixture, the whole density is described by
the spinless-fermions solution. Thus, the whole density follows the
boomerang dynamics as expected for non-interacting fermions or
Tonks-Girardeau bosons \cite{Janarek2023}.  The effect of the
interactions manifests itself in the inter spin-component dynamics,
but in a very different way depending on whether the interactions are
infinite or finite.  When the interactions are infinite, each spin
component is just a system of non-interacting particles that is not
free to propagate everywhere because of the presence of the other
component.  It localizes at the same time as the whole system, but
largely far away from the position of the center-of-mass of the whole
system and also largely far away from their initial position. The
center-of-mass of each component does not do the boomerang dynamics:
each component moves away but does not come back because it is pushed
by the other component.

Instead, when the interaction is large but finite, the situation is
completely different.  The landscape of the disorder felt by the spins
completely change. Particles with different spin hop with a spatially
random probability, that fluctuates as a function of time, and is
inversely proportional to the interaction strength.  The
center-of-mass of each component reaches a final position that is
different from the initial one, as already found for other interacting
systems \cite{Janarek2023}, and, coincides with that of the whole
system.

From the point of view of each component, there is an initial time
when the dynamics at finite interactions coincides with that at
infinite interactions, but then the trajectories separate, one going
back while the other does not.  This is very different from what
happens with a single-component system \cite{Janarek2023} where the
center-of-mass for the gas of finite interactions slightly deviates
from that with infinite interactions.

As a final remark, we would like to highlight the role of symmetries
for the QBE. As pointed out in \cite{Noronha2022,Janarek2022}, the QBE
takes place in real space if the ensemble of the disordered
Hamiltonians $\{\mathcal{H}\}$ is invariant under the action of
$\mathcal R \mathcal T$ and if the initial state is an eigenstate of
$\mathcal R \mathcal T$, where $\mathcal R$ is the spatial reversal
operator and $\mathcal T$ the time-reversal operator.  For the system
we have considered in this work, both conditions are fulfilled.
However, each spin component {\it is not} an eigenstate of
$\mathcal R \mathcal T$. Under the action of $\mathcal R \mathcal T$
the spin-up component on the left is transformed on the spin-down
component on the right, and viceversa.  Thus, each spin-component
separately does not fulfil the condition for the QBE, but the two
spin-components system does.

\section{The thermalization issue}
\label{sec-therm}
\begin{figure}
    \centering
  \includegraphics[width=0.9\columnwidth]{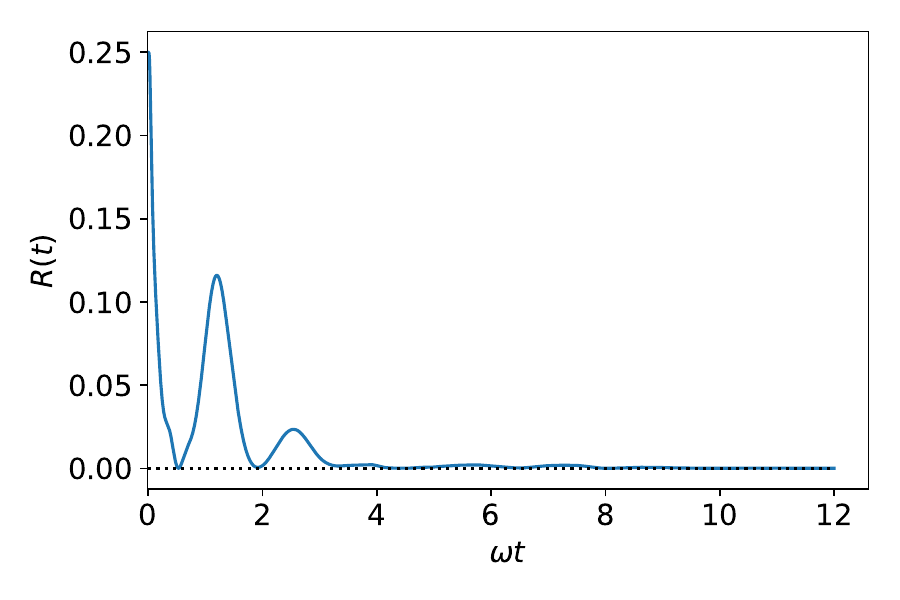}
    \caption{Distance $R(t)$ from the spin part of the time-dependent average spin density $\bar{\rho}_{\uparrow}(t)$ to the microcanonical one $\rho_{\uparrow,\text{MC}}$
     as function of time $t$.
     \label{fig-rt}}
\end{figure}

Given that the system we are studying is characterized by interactions
and disorder, it is natural to ask whether the final localized state
is many-body localized or if the eigenstate thermalization hypothesis
(ETH) holds. In the latter case, if the expectation value of local
operators will ultimately evolve in time to their value predicted by
the microcanonical ensemble \cite{Alet2018,pecci2022}.  Such a target
state, for our system, coincides with a state described by the
diagonal ensemble where the spin densities
$\rho_{\uparrow,\text{MC}}(x)$ and $\rho_{\downarrow,\text{MC}}(x)$
are equally distributed.  Of course, this cannot happen at infinite
interactions: in this limit the system is integrable and the different
spins, that are initially spatially phase separated, never mix, making
impossible to achieve a uniform non-magnetized state.  But at large
finite interactions, even if slowly, the spin mixing takes place and
continues even after the total density has localized. The long-time
position of the center-of-mass of each spin component coincides with
the position of the center-of-mass of the whole localized cloud. This
is in accordance with what one would expect with the aforesaid fully
mixed state where
$\rho_{\uparrow,\text{MC}}(x)=\rho_{\downarrow,\text{MC}}(x)$.
Furthermore, since we have verified that the particle densities
$\rho^i(x,t)$ and amplitudes $|[a_P(t)]_i|^2$ entering the spin
densities (cf. Eq. (\ref{eq:rhodef})) possess different relaxation
times, in order to investigate the thermalization of the spin dynamics
we have evaluated the distance
\begin{equation}
R(t)=\sum_{i}  \left(\overline{\sum_{P \in S_N}|[a_P(t)]_i|^2}-\sum_{P \in S_N}|[a_P]^\text{MC}_i|^2\right)
\label{eq:Rt}
\end{equation}
where $[a_P]^\text{MC}_i$ are the coefficients obtained from the
diagonal ensemble, analogously to what has been done in
\cite{pecci2022} (see Fig. \ref{fig-rt}). $R(t)$ collapsing to zero
very rapidly, it is clear that, for what concerns the spin density
distributions, our system verifies ETH.  Thermalization was also
reported for spin chains subject to off-diagonal disorder
\cite{Luitz2015}. Our case displays some differences with the above,
as the effective disorder felt by the spins is time-dependent, as it
originates from the dynamics of the orbital part. It is also
interesting to mention that the disorder felt by the particles is
diagonal, but it turns off-diagonal as felt by the spins because it is
mediated by the interaction among particles.

In conclusion, we would like to emphasize that the center-of-mass
evolution of the spin components, namely whether the boomerang
dynamics occurs for the spin components, is an experimentally
accessible tool to probe ETH or lack of thermalization.

\section{Concluding remarks}
\label{sec-concl}
In this work we have studied the dynamics of two-component fermionic
wavepacket launched in a 1D pseudo-random potential. We have
considered the case of an initially spatially phase-separated fermions
characterized by a strong repulsive inter-species contact
interactions.  In such a strongly interacting limit, the charge and
the spins dynamics decouple.  The total density coincides with the
total density of a non-interacting spinless Fermi gas, while the spin
components obey to an effective non-homogeneous Heisenberg spin chain
Hamiltonian, whose hopping terms, that depend on the density
evolution, become random during the dynamics and fluctuate in time.
As a result, we find that the total density performs a boomerang
dynamics as predicted for non-interacting particles, while the
densities of each spin components, considered separately, do not.
Their centers of mass initially move away and then they come back, not
to their respective initial position, but towards the initial position
of the whole system. They reach this position, making damped
oscillations, whose frequency is determined by averages of the
effective hopping energy.  The two spin-components mixed together
reach a final spin-density distribution that is compatible with that
of the diagonal (microcanonical) ensemble. This is a signature that
interactions, in our system, do not induce many-body localization, at
least for the parameters that we have chosen for our study. This
result is reminiscent of previous studies \cite{Alet2018} that have
shown that for an Heisenberg spin chain with off-diagonal quenched
disorder, many-body localization does not occur.  Finally, let us
underline that the system we have studied in this work opens up the
possibility to realize, with an ultracold atom experiment, a quantum
simulator of a Heisenberg spin chain with off-diagonal disorder.
However, our system differs from that analyzed in
\cite{Laflorencie2003,Laflorencie2003b,Laflorencie2004,Laflorencie2005}
as the spatial part is subject to a diagonal disorder and the spin
components feel an off-diagonal time-dependent disorder. This could
bring unexpected novel phases and deserve further studies for larger
systems, correlated disorder and finite temperature.

\section*{Acknowledgments}
The authors acknowledge the CNRS International Research Project COQSYS
for their support, funding from the ANR-21-CE47-0009 Quantum-SOPHA
project, support of the Institut Henri Poincar\'e (UAR 839
CNRS-Sorbonne Universit\'e), and LabEx CARMIN (ANR-10-LABX-59-01), and
support from the Scientific Research Fund of Piri Reis University
under Project Number BAP-2024-04. P. C. acknowledges support from
CONICET and Universidad de Buenos Aires, through grants PIP
11220210100821CO and UBACyT 20020220100069BA, respectively.
L.T. acknowledges the financial support of the CIC-UMSNH grant
n. 18090.  The authors acknowledge useful discussions with Dominique
Delande, Thierry Giamarchi and Nicolas Laflorencie. They wish to
dedicate this paper to Dominique Delande who passed away during the
progress of this work.

\appendix
\section{Two spin dynamics}
\label{app1}
\begin{figure}
  \includegraphics[width=0.8\linewidth]{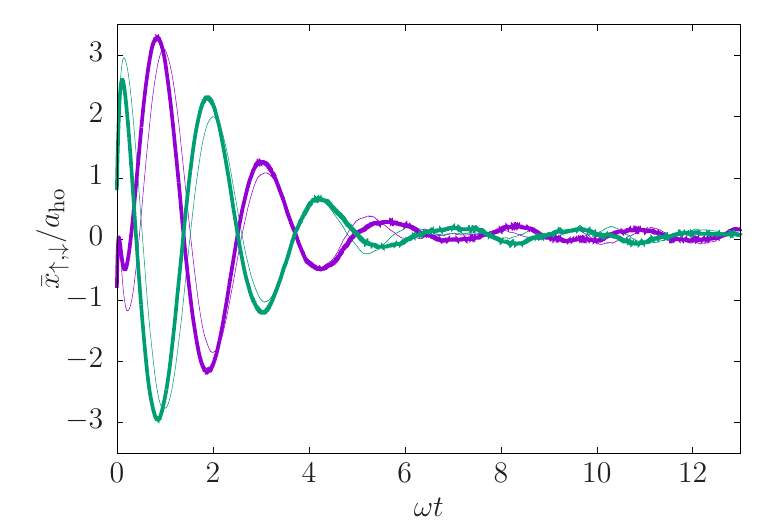}
  \caption{\label{fig-app} Spin-up (violet curves) and spin-down
    (green curves) center-of-mass positions $\bar x_\uparrow$ and
    $\bar x_\downarrow$ in units of $a_{ho}$ as functions of the time
    $t$, for the case of a two particles systems. The exact results
    (thick lines) are compared with those obtained using
    Eq. (\ref{approx-2part}).}
\end{figure}

Let us consider the case of two fermions.  In this case, the spin part
of the many-body wavafunction can be written on the snippet basis as
$a_1(t)|\uparrow\downarrow\rangle+a_2(t)|\downarrow\uparrow\rangle$,
with $a_1(t=0)=1$ and $a_2(t=0)=0$.  There exists only a hopping term
$J$ and the $a_i$'s obey the differential coupled equations
\begin{equation}
    i\hbar\dot a_{1,2}=-J a_{1,2}+ J a_{2,1}.
\end{equation}
Taking into account that $a_1^2+a_2^2=1$,
then we get
\begin{equation}
    \begin{split}
        a_1^2&=\dfrac{1}{2}\left(1+\cos(2\int_0^t J(t') dt')\right)\\
        a_2^2&=\dfrac{1}{2}\left(1-\cos(2\int_0^t J(t') dt')\right)\\
    \end{split}
\end{equation}
The center-of-mass of each component can be written
\begin{equation}
\begin{split}
    \bar{x}_{\uparrow}&= \overline{a_1^2 x_1+a_2^2 x_2}\\
    \bar{x}_{\downarrow}&= \overline{a_2^2 x_1+a_1^2 x_2}
    \end{split}
\end{equation}
where $x_1$ and $x_2$ are the baricenters of the density distributions
$\rho_1$ and $\rho_2$.
We observe that
\begin{equation}
\begin{split}
    \bar{x}_{\uparrow}&\simeq \overline{a_1^2} \bar x_1+\overline{a_2^2}\bar x_2\\
    \bar{x}_{\downarrow}&\simeq \overline{a_2^2}\bar x_1+\overline{a_1^2}\bar x_2.
    \end{split}
\end{equation}
Since in the two particles case $\bar x_1=\bar x_\uparrow^\infty$ and  $\bar x_2=\bar x_\downarrow^\infty$,
we obtain
\begin{equation}
    \bar{x}_{\uparrow,\downarrow}=\dfrac{1}{2}(\bar x_\uparrow^\infty+\bar x_\downarrow^\infty)
    \pm\dfrac{1}{2}(\bar x_\uparrow^\infty-\bar x_\downarrow^\infty)\overline {\cos(2\int_0^t J(t') dt')}.
    \label{approx-2part}
\end{equation}
The $J$'s distribution not being gaussian, the disorder average of the
cosine function cannot be written in simple terms, and we needed to
compute it numerically.  Indeed even if $\bar J$ determines the spin
oscillation frequency, we have verified that the variance $\sigma_j$
does not allow to deduce a correct damping time.  The comparison
between the center-of-mass exact evolution and the approximated one
given in Eq. (\ref{approx-2part}) is shown in Fig. \ref{fig-app}.

\bibliography{biblionew}
\end{document}